\documentclass[aps,amsfonts,amsmath,prd,preprint,nofootinbib]{revtex4-1}
\pdfoutput=1
\newcommand{\beq}{\begin{equation}}
\newcommand{\eeq}{\end{equation}}
\usepackage{epsfig}

\def\be{\begin{equation}}
\def\ee{\end{equation}}
\def\beqa{\begin{eqnarray}}
\def\eeqa{\end{eqnarray}}

\begin{document}

\title{Phenomenology of the Watcher measure in the Bousso-Polchinski landscape}

\author{Delia Schwartz-Perlov}

\affiliation{
Institute of Cosmology, Department of Physics and Astronomy,\\ 
Tufts University, Medford, MA 02155, USA
}

\begin{abstract}

We investigate the phenomenology of the ``Watcher'' measure in the Bousso-Polchinski landscape.  We study a small toy landscape analytically. The results are sufficient to allow us to extrapolate ``watcher'' phenomenology to much larger landscapes.  We compare our results to other measures which have been applied to BP landscapes containing googles of vacua.  Under certain well motivated conditions, watcher phenomenology coincides with the causal patch prescription, but not Linde's volume weighted measure.  
\end{abstract}

\maketitle

\section{Introduction}

The emerging picture from eternal inflation and string theory is that we live in a multiverse which consists of an unlimited number of bubble universes, or vacuum states, that come in a vast, but finite, number of types.  Different vacuum states in the string theory landscape have their own particles, interactions and constants of nature.  Moreover, they are metastable, like the false vacuum of inflation.   Thus, if the universe begins in some initial vacuum state, all other vacua can be realized via ongoing bubble nucleation.  In this way the multiverse can be populated with each type of universe permitted by the string theory landscape.

If this picture is correct, then we can no longer predict with certainty what the value of a constant that varies in the landscape is (for example, the cosmological constant).  But we can, \textit{in principle},  calculate how the constants of nature are distributed amongst all the different bubble types.  Figuring out how to do so, in practice, is called the measure problem. But why is it a ``problem''?

Let us define the relative probability for the observation of two outcomes A and B to be the ratio

\beq
{{P(O_A)}\over{P(O_B)}}={{N_A}\over{N_B}},
\eeq

where $N_i$ is the number of times an observation of type $i$ is measured.

It turns out that in an eternally inflating universe the number of each type of vacuum
diverges as $t\to\infty$.  Thus any outcome, which is not strictly forbidden, will occur an infinite number of times, and so all $N_i$ are infinite.  

One of the most fruitful approaches to taming these infinities has been to impose a global time cut-off \cite{Linde:1993nz,Linde:1993xx,GarciaBellido:1993wn,Vilenkin:1994ua,DeSimone:2008bq,Bousso:2008hz,DeSimone:2008if,Garriga:2005av,Bousso:2009dm,Vilenkin:2011yx,Salem:2011mj,Linde:2008xf}.  The basic idea is as follows:  Choose an initial spacelike hypersurface $\Sigma_0$. Then consider a congruence of future-directed timelike geodesics.
These geodesics sweep out a spacetime volume between the initial hypersurface and some final hypersurface  $\Sigma_{t}$.   
Count the number of different types of events within the finite spacetime region between these two hypersurfaces, and then take the limit $t \rightarrow \infty$ to get the relative probability of any two events $ {{P(O_i)}\over{P(O_j)}}=lim_{t \rightarrow \infty} {{N_i(t)}\over{N_j(t)}}$.  Unfortunately, for these so-called ``global'' measures, one finds very different results depending on the choice of the time variable $t$ \cite{Linde:1993nz,Linde:1993xx,GarciaBellido:1993wn}. (For more recent discussions, see \cite{VVW,Guth00,Tegmark}.)  


In addition to the proposed global measures, several ``local'' measures, which sample the space-time region around a chosen timelike geodesic, have also been proposed \cite{Bousso:2006ev,Bousso:2010zi,Garriga:1997ef,Vanchurin:2006qp,Vanchurin:2006xp,Nomura:2011dt}. One can count observations made by all observers in some (non-unique) space-time region around the geodesic; or one can consider only the observations of a single observer, defined by a timelike geodesic . For an up-to date review of the measures on the market see \cite{Freivogel:2011eg}.  In this paper, we will focus on the phenomenology of the recently proposed ``watcher''  measure \cite{Watchers}. 

The watcher measure falls under the category of a single observer local measure.  In the simplest single observer measures \cite{Garriga:1997ef,Vanchurin:2006qp,Vanchurin:2006xp}, a typical geodesic is chosen to start in some initial metastable dS vacuum state.  The geodesic will then pass through a finite sequence of dS vacua, before it hits a terminal AdS vacuum, which will undergo a big crunch; or until it hits a terminal supersymmetric stable Minkowski vacuum.   One can determine the probability for the world line to be in a given vacuum $j$, by counting the frequency with which the world line enters vacuum $j$.  Unfortunately, the results are sensitively dependent on which geodesic is chosen.  Thus, one needs to choose an ensemble of geodesics, with different initial conditions.  There is currently no ``correct'' way to do so.

In the watcher measure \cite{Watchers}, Garriga and Vilenkin make the non-standard assumption that AdS crunches do not result in a singularity.  Rather, they propose that a crunching vacuum will undergo transitions to other vacuum states in the landscape.  Thus, in this picture, a typical geodesic will undergo an infinite sequence of visits to all the dS and AdS vacua in the landscape\footnote{For our purposes here, we will only consider irreducible landscapes with dS and AdS vacua - we will not consider Minkowski vacua.  The interested reader can refer to \cite{Watchers} to see how the watcher measure can incorporate Minkowski vacua.}. This future-eternal time-like geodesic can be thought of as the world-line of a ``watcher'', who samples different events as the history of the multiverse unfolds.  Any event which has a non-zero probability to occur, will do so an infinite number of times \cite{Garriga:2001ch}.  Thus the relative probability, $P_A/P_B$, of two events, A and B, can be defined as the relative frequency with which the events are encountered by the watcher's geodesic.  In particular \cite{Watchers},
\beq
P_A \propto \sum_j X_j N_A^{(j)}
\eeq

where $X_j$ is defined as the frequency with which the watcher visits vacuum $j$, $N_A^{(j)}$ is the average number of type A events encountered during a visit to vacuum $j$, and the sum is over all vacua.  The average number of events of a given type occurring in a vacuum, $N_A^{(j)}$, depend on the physics of that vacuum.  We will only focus on calculating the frequencies, $X_j$, for which there is a well-defined formalism which we will review in section \ref{frequencieswatchermeasure}.   In addition to other desirable features, like avoiding Boltzmann Brain dominance, this model is initial condition independent.


We will apply the watcher measure \cite{Watchers} to a toy model of the string theory landscape developed by Bousso and Polchinski \cite{BP} (hereafter BP).
In the BP model, $J$ different four-form fluxes, $F_a$, give rise to a $J$-dimensional grid of vacua, each
labeled by a set of integers $n_a$, which represent the number of units of a particular flux\footnote{In general, the $n_a$ can be positive and negative.}. Each point in the BP grid corresponds
to a vacuum with the flux values $F_a = n_a q_a$ and a
cosmological constant
\be
\label{totalLambda}
\Lambda=\Lambda_{bare}+\frac{1}{2}\sum_{a=1}^{J}F_a^2=\Lambda_{bare}
+\frac{1}{2}\sum_{a=1}^{J}n_a^2q_a^2.
\ee

$\Lambda_{bare}$ is the bare cosmological constant, which is assumed to be large and negative ($|\Lambda_{bare}|\sim 1$); $q_a$ is the charge associated with the corresponding four-form flux $F_a$, and is assumed to be a non-small value ($q_a \sim 0.1$).  
The positive-energy vacua of the BP grid will undergo exponential inflationary expansion. The higher the energy, the greater is the expansion rate.  The flux configuration in inflating regions can change from one point on the grid to the next
through bubble nucleation. In particular, the change of the field strength across a nucleated bubble is
\be
\Delta F_a =\pm q_a,
\label{DeltaF}
\ee

and the corresponding change in $\Lambda$ is

\beq
|\Delta\Lambda_a|=(n_a\pm1/2)q_a^2.
\eeq

Thus, bubbles nucleate within bubbles, and each time this happens the cosmological constant either increases or decreases by a discrete amount. This mechanism allows the universe to
start off with an arbitrary cosmological constant, and then to diffuse through the BP grid of possible vacua, to generate regions with each and every possible cosmological constant.  BP showed that with $J\sim 100$, the spectrum of allowed values of $\Lambda$ can be dense enough to include vacua which have a tiny cosmological constant (like we observe in our universe), thereby offering a resolution to the cosmological constant problem.

The first measure prescription to be applied to the BP landscape was the comoving horizon cutoff (CHC) method or the ``bubble abundance'' prescription \cite{Garriga:2005av,SchwartzPerlov:2006hi}.  This was followed by Bousso's causal patch method \cite{Bousso:2006ev,Bousso:2007er}, and then by Linde's volume weighted measure \cite{Linde:2007nm,Clifton:2007bn}.  We compare the results of the watcher prescription with these studies.  We find that the watcher measure favors  the dynamical selection of low-energy vacua with ``diagonal'' flux configurations (this means that all of the $J$ fluxes have quanta that are approximately equal), like the bubble abundance and causal patch measures.  Volume weighted measures favor ``axial'' flux configurations. 

The outline of this paper is as follows:  In Section \ref{frequencieswatchermeasure} we summarize the prescription of Ref.~\cite{Watchers} for calculating the frequencies with which the watcher visits different vacua, $X_j$.  In Section \ref{BPtoymodel} we apply the ``watcher'' prescription to calculate probabilities analytically for a small toy BP model.  We then compare the predictions of the watcher measure to that of the three other measures which have been applied to the BP landscape in Section \ref{comparison}. Conclusions follow in Section \ref{conclusions}.  The calculation of the transition rates for the BP model is reviewed in the Appendix.    

\section{Frequency of visits in the watcher measure}\label{frequencieswatchermeasure}

In this section we present the main relevant results of \cite{Watchers}, and closely follow the notation developed therein.  Consider the evolution of an ensemble of watchers in a non-terminal irreducible\footnote{In an irreducible landscape, all vacua can be reached from any others by some sequence of decays.} landscape.  Such a landscape has only ``recyclable'' vacua; either the vacua are dS states with positive energy density ($\Lambda>0$), or if they are AdS states with negative energy density ($\Lambda<0$), then they can tunnel to dS states after they undergo a crunch.  The watchers evolve independently, even though they are statistically the same.  

To calculate the frequency with which a watcher visits different vacua, a discrete time variable $n$, which increases by one whenever a watcher makes a transition to a new vacuum state, is introduced.  The fraction of watchers in vacuum $J$ at the time step $n$, $X_J(n)$, obeys the evolution equation

\beq
X_I(n+1)=\sum_J T_{IJ} X_J(n), \label{evolutionequx}
\eeq

where the transition probabilities satisfy
\beq
\sum_I  T_{IJ} =1.
\eeq
Notice that $T_{II}=0$ because the watcher is required to jump to a new vacuum at every time step.

Capital indices label both dS and AdS vacua.  If we label dS vacua with indices from the middle of the alphabet, and AdS vacua with indices from the beginning of the alphabet, then Eq. (\ref{evolutionequx}) can be rewritten as:

\beq
X_i(n+1)=\sum_j T_{ij}X_j(n)+\sum_a T_{ia}X_a(n). \label{dSevolutionequx}
\eeq

and

\beq
X_a(n+1)=\sum_j T_{aj}X_j(n)+\sum_b T_{ab}X_b(n). \label{AdSevolutionequx}
\eeq

For transitions from dS vacua, the transition matrix is defined as 
\beq
T_{Ij} \equiv {{\kappa_{Ij}} \over {\kappa_j}}.
\eeq

with 

\beq
\kappa_j \equiv \sum_R \kappa_{Rj}.
\eeq

The transition rate $\kappa_{Ij}$ is defined as the probability per
unit time for a watcher who is currently in vacuum $j$ to find
herself in vacuum $I$.  Its magnitude depends on the choice of the time variable $t$. Choosing the proper time $t$,
\beq
\kappa_{Ij}=\Gamma_{Ij}{4\pi\over{3}}H_j^{-3},
\label{kappa}
\eeq
where $\Gamma_{Ij}$ is the bubble nucleation rate per unit
physical spacetime volume 
and
\beq
H_j = (\Lambda_j/3)^{1/2}
\label{Hj}
\eeq
is the expansion rate in vacuum $j$.\footnote{Here and below we use the reduced
Planck units, $M_p^2/8\pi = $ where $M_p$ is the Planck mass.}  

In the following section we will make some operational assumptions about the transition probabilities (also called branching ratios) from AdS vacua $T_{Ia}$. 

In \cite{Watchers} it was shown that there exists an asymptotic stationary distribution for $X_J(n)$ which can be found by solving the equation,

\beq
\sum_J T_{IJ} X_J=X_I \label{asympdistrib}.
\eeq

Let us now solve this equation for a toy landscape.

\section{Bousso-Polchinski toy model} \label{BPtoymodel}

\begin{figure}
\centering
\includegraphics[width=5.0in]{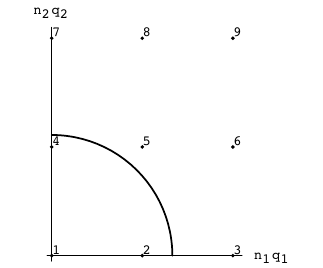}
\caption{The arrangement of vacua for a $J=2$ BP
grid with $0 \le n_a \le 2$.} \label{2DanalyticBP}
\end{figure}

We will apply the watcher prescription to calculate frequencies for a very
simple BP model, which can be solved analytically. We consider $9$
vacua arranged in a 2-D grid and labeled as indicated in Fig.
\ref{2DanalyticBP}. There are three AdS vacua labeled 1, 2, 4,
and six dS vacua, 3, 5, 6, 7, 8, 9 in this model. We allow
upward and downward transitions between nearest neighbor dS pairs, and downward
transitions from dS to AdS states.   In addition, we allow transitions from crunching AdS states to the dS vacua.  We will assume that the transition probabilities from AdS vacua to dS vacua are independent of the crunching AdS vacuum $a$, thus
\beq
T_{ja} \equiv Q_j.
\eeq

We also assume that there are no transitions between two AdS vacua, so  $T_{ab}=0$.


Under these assumptions, equation (\ref{asympdistrib}) can be written as

\beq
\sum_j \left(T_{ij}-\delta_{ij}\right)X_j=-\xi Q_i
\eeq

with
\beq
\xi \equiv \sum_a X_a.
\eeq

For our toy landscape, this is explicitly given by

\be
\left(%
\begin{array}{cccccc}
  -1 & 0 &\kappa_{36}/\kappa_6& 0 & 0 & 0 \\
  0 &-1& \kappa_{56}/\kappa_6 & 0 & \kappa_{58}/\kappa_8& 0 \\
\kappa_{63}/\kappa_3& \kappa_{65}/\kappa_5& -1& 0 & 0& \kappa_{69}/\kappa_9 \\
  0 & 0 & 0&-1 & \kappa_{78}/\kappa_8& 0 \\
    0 & \kappa_{85}/\kappa_5 & 0&  \kappa_{87}/\kappa_7& -1 & \kappa_{89}/\kappa_9\\
      0 & 0 & \kappa_{96}/\kappa_6& 0& \kappa_{98}/\kappa_8&-1 \\
\end{array}%
\right)\left(%
\begin{array}{c}
  X_3 \\
  X_5 \\
 X_6\\
 X_7 \\
 X_8\\
 X_9 \\
\end{array}%
\right)
=-\xi\left(%
\begin{array}{c}
  Q_3 \\
   Q_5 \\
 Q_6\\
 Q_7 \\
  Q_8\\
 Q_9 \\
\end{array}%
\right)
\label{matrixinvert}
\ee

with
\be
\xi=X_1+X_2+X_4.
\eeq

This system of equations can be solved exactly by inverting the matrix on the left hand side.  However, we notice that the upward transitions in the numerators of the lower diagonal components are exponentially suppressed with respect to the components in the upper diagonal part of the matrix.  Thus, to very good approximation\footnote{Note that we are implicitly assuming that all vacua in our model are well below the Planck regime.  However, in large generic landscape models, we expect that as we approach the Planck regime, upward transitions become less and less suppressed, and can become comparable to downward transitions.} we consider instead the simpler set of equations

  \be
\left(%
\begin{array}{cccccc}
  1 & 0 &-\kappa_{36}/\kappa_6& 0 & 0 & 0 \\
  0 &1& -\kappa_{56}/\kappa_6 & 0 &-\kappa_{58}/\kappa_8& 0 \\
0& 0& 1& 0 & 0& -\kappa_{69}/\kappa_9 \\
  0 & 0 & 0&1 & -\kappa_{78}/\kappa_8& 0 \\
    0 & 0 & 0& 0& 1 & -\kappa_{89}/\kappa_9\\
      0 & 0 & 0& 0& 0&1 \\
\end{array}%
\right)\left(%
\begin{array}{c}
  X_3 \\
  X_5 \\
 X_6\\
 X_7 \\
 X_8\\
 X_9 \\
\end{array}%
\right)
=\xi\left(%
\begin{array}{c}
  Q_3 \\
   Q_5 \\
 Q_6\\
 Q_7 \\
  Q_8\\
 Q_9 \\
\end{array}%
\right)
\label{matrixinvert}
\ee

The solution\footnote{All of the $X_i$'s are proportional to $\xi$, so this constant factor will cancel out when we consider the ratio of any two frequencies.} for the frequency of each vacuum is:
\beq
X_9=\xi Q_9
\eeq

\beq
 X_8=\xi Q_8+{{\kappa_{89}}\over{\kappa_9}}X_9
\eeq

 \beq
 X_7=\xi Q_7+{{\kappa_{78}}\over{\kappa_8}}X_8
 \eeq  
  
\beq
  X_5=\xi Q_5+{{\kappa_{56}}\over{\kappa_6}}X_6+{{\kappa_{58}}\over{\kappa_8}}X_8
\eeq  

 \beq
 X_6=\xi Q_6+{{\kappa_{69}}\over{\kappa_9}}X_9
\eeq  

  \beq
  X_3=\xi Q_3+{{\kappa_{36}}\over{\kappa_6}}X_6 \label{directpath3}
\eeq

By inspection we can generalize the solution, so that  
\beq
X_i=\xi Q_i+\sum_j{{\kappa_{ij}}\over{\kappa_j}}X_j \label{downwardsol}
\eeq
where $j$ labels all the parent vacua that can nucleate type $i$ bubbles.  It is important to note that here type $j$ vacua have higher energy than type $i$ vacua.  Also,  types $j$ and $i$ vacua are dS vacua.
  
Notice that since vacuum 9 is the highest energy vacuum in this toy landscape, it cannot be reached from a parent vacuum.  Also, vacuum 5 can be reached from two parent vacua, thus there are two terms in the sum over branching ratios.

If we do not ignore the upward tunneling matrix elements, then in addition to the paths that lead to a given vacuum via a downward cascade, there are closed (or virtual) ``probability currents'' which also contribute, but they are suppressed by at least two branching ratios (one of these includes an exponential suppression for an upward jump).  An example of a closed probability current in our toy model follows: Let's assume we are trying to find $X_3$. There is a downward path from vacuum $9$ to $6$ and then $3$.  This contribution is captured by Eq. (\ref{directpath3}).  But if all upward transitions are included, then there will also be a contribution to $X_3$ of the form $\left(1-{{\kappa_{87}}\over{\kappa_7}}{{\kappa_{78}}\over{\kappa_8}}+ \cdots \right) \times X_{3}^{down}$, where $X_{3}^{down}$ is the frequency with only downward jumps (the $+ \cdots $ are similar terms which involve other dS vacua which are not part of the direct downward chain to vacuum $3$). We will not consider these small ``virtual'' contributions in the rest of the paper.

\subsection{Choosing the $Q_i$'s}
As discussed in Ref. \cite{Watchers}, there is currently no unique way to calculate what the $Q_i$ should be.  So we will explore three different options here.

Firstly, let us begin by assuming that $Q_9=1$, and all other $Q_i$ are zero, then we see that the watcher measure will predict the following:

\beq
X_5 \approx 2 X_7 \approx 2 X_3
\eeq

Here we have assumed that all branching ratios are approximately $1/J$, which is $1/2$ in this $2$ dimensional model.  Since vacuum 5 can be reached via two parent vacua, it has twice the probability compared to the other two penultimate\footnote{Penultimate vacua are dS states which have all their downward decay products in the AdS regime.} vacua in this toy model.  One can calculate $X_8$ and $X_6$ too, but we will ignore them because they are not penultimate vacua.  In a larger landscape we would only be interested in the distribution of penultimate vacua, as we expect any habitable vacua to be of this sort. 

Now we will assume that all $Q_i$ are equal, thus $Q_i= 1/N=1/6$ where $N$ is the number of dS vacua. Let us continue to assume that all the branching ratios are approximately $1/J$. Then, for our toy model, we find

\beq
X_i=\xi Q_i(1+n{{1-J^{-m}}\over{J-1}} )\label{equalQequalbranch}
\eeq

where $n$ is the number of parent vacua for vacuum $i$, and we have used

\beq
\sum_{s=1}^m {{1}\over{J^S}}={{1-J^{-m}}\over{J-1}}
\eeq
 
where $m$ is the number of steps $s$ taken to get from the highest vacuum (in this case vacuum $9$) to vacuum $i$.  Here $m=2$.  Thus we find,
\beq
{{X_5}\over{X_7}} ={{X_5}\over{X_3}}=10/7.
\eeq
Thus, although $X_5$ is still greater than $X_7$ and $X_3$, the distribution is somewhat ``smoothed" compared to the first scenario where we assume $Q_9=1$, and other $Q_i=0$.

Finally, we will look at the special case of $Q_5=1$, and all others are zero.  If we ignore upward jumps, as we did to get the solution in Eq. (\ref{downwardsol}), then the watcher will only visit a type 5 vacuum ($X_5=1$, all other $X_i=0$).  But, if we include upward jumps, then we find that the frequencies are given roughly by the product of branching ratios along the path from the starting vacuum to the vacuum in question.  This situation is reminiscent of the bubble abundance measure \cite{SchwartzPerlov:2006hi} which picks the slowest decaying state, the dominant vacuum, to be in essence the initial condition\footnote{The dominant vacuum is a de Sitter vacuum, with small, but not necessarily tiny, $\Lambda$.  It was shown in \cite{SchwartzPerlov:2006hi} that all downward jumps from the dominant vacuum lead to AdS vacua.}.  All other vacua reached from the dominant vacuum acquire non-zero probabilities through a series of upward and downward cascades.  Vacua which are closest in configuration space to the dominant vacuum will have the highest probabilities.  It was illustrated in \cite{SchwartzPerlov:2006hi} that we expect the dominant vacuum to have a flux configuration with all flux quanta approximately equal.  This is because the decay rates are most sensitive to the flux quantum numbers $n_a$ (the size of the charge $q_a$ also enters; see the Appendix for more details of the decay rates in the BP model ), and if there is one quantum which is much bigger than the others, the decay will be quite rapid in that direction, and thus we would not expect the corresponding vacuum to be the slowest decaying state.  Thus we predict, for this particular choice of $Q_i$, that the most frequented low energy vacua would have flux coordinates that lie close to the diagonal in configuration space which is closest to the dominant vacuum.

 \section{Generalizing to a larger BP landscape and comparison with other measures} \label{comparison}
 
Although we applied the watcher measure to a very simple toy model, we can readily generalize the results to a much larger BP landscape.  Of particular interest is how watcher measure phenomenology compares to that of other measures which have already been applied to large BP landscapes.

\subsection{Bubble abundance prescription}
The first prescription to be applied to the BP landscape was  the comoving horizon cutoff (CHC) method or the ``bubble abundance'' prescription \cite{Garriga:2005av,SchwartzPerlov:2006hi}.  As touched on in the previous section, the bubble abundance prescription (although it is initial condition independent),  singles out the ``dominant vacuum'' as a starting point for the further evolution of the flow of probability through the BP landscape.  Vacua reached via the least amount of steps, and with the highest product of branching ratios from the dominant vacuum, are the most probable.  The dominant vacuum should have a flux configuration close to a diagonal on the grid\footnote{In the BP landscape, there will be $2^J$ diagonals, if flux quantum numbers are allowed to be positive and negative.  In this case, the dominant vacuum will be a penultimate vacuum that lies close to one of the many diagonals. If we consider only positive flux quantum numbers then there will be only one diagonal in the BP grid.}.  Thus, we expect vacua with similar flux configurations to be more abundant than other vacua which are further away in flux space.  
In the watcher measure, if all crunching vacua transition only to the slowest decaying dS vacuum ($Q_{dom}=1$  - this corresponds to $Q_5=1$ in the toy model), then we would find the same qualitative result as in the bubble abundance case; namely vacua with flux configurations close to that of the dominant vacuum should be visited more frequently.  Exact ratios of frequencies depend on the detailed parameters of the BP model.  Of course, $Q_{dom}=1$, and all other $Q_i=0$ is an unlikely and unmotivated scenario.  However, we consider it just to elucidate the relationship to the bubble abundance results.

\subsection{Causal patch probability prescription}
In  \cite{Bousso:2006ev,Bousso:2007er}, Bousso and collaborators developed and applied the ``local'' or ``causal patch'' measure to a large BP landscape, having googles of vacua.   They considered an ensemble of worldlines, each starting in a vacuum randomly drawn from an initial distribution (their measure is initial condition dependent).  Each worldline passes through a sequence of metastable dS vacua.  At each step in the decay chain, a new vacuum $i$ is entered with a probability given by the transition probability $T_{ij}$, where $j$ is the parent vacuum.  The downward cascade from high to low energy vacua, terminates when the worldline enters an AdS vacuum.  This measure defines the probability of vacuum $i$ to be given by the product of transition probabilities from each initial vacuum which can reach the vacuum $i$,  along the shortest possible paths, weighted by the initial probability distribution. 

Using Monte Carlo simulations, Bousso et al  \cite{Bousso:2007er} computed the probability distribution of the low energy, penultimate vacua.  They chose an initial probability distribution  $P_i^{(0)}$ which depends only on $\Lambda$, and favors high energy vacua (this type of initial condition can be motivated by the tunneling wave function \cite{Vilenkin:1984wp,Linde:1983mx} of quantum cosmology).  They found that low energy vacua with diagonal flux configurations are preferentially selected during the cosmological dynamics. 

This type of initial condition is analogous to our toy model, where we chose $Q_9=1$.  We expect that if the watcher measure were applied to the same BP grid that Bousso et al used, with the $Q_i$ chosen in the same way as their initial probability distribution, we would make the same predictions.  So let us briefly discuss why, under these conditions, low energy ``diagonal'' vacua are preferentially selected in watcher and causal patch phenomenology.

Let us imagine that there is a Plank scale, high energy shell in the BP grid, above which all vacua have the same $Q_i$ or $P_i^{(0)}$, for the watcher and causal patch measures, respectively.  Below this shell, we shall assume all $Q_i=0$ and  $P_i^{(0)}=0$.  Now, if we follow a world line that starts close to an axis in flux space, the decay chain will follow a sequence which will favor jumps along the direction of that axis (because jumps in a  high flux quantum direction have the highest tunneling rates, and thus the transition probabilities are maximized.)  If, instead, we follow a world line that has an initial flux configuration close to a diagonal, then the sequence of decays will ``zig-zag'' down the diagonal - once again, because decays in directions with high flux quanta occur with higher probability.  Worldlines which start off somewhat displaced from the diagonal, will first tunnel ``straight down'' in the largest flux quanta directions, until they enter vacua with a diagonal flux configuration.  They will then ``zig-zag'' down the rest of the diagonal.  The transition probabilities are smaller along ``diagonal'' paths than ``axial'' paths  - but not exponentially smaller.  We expect axial transition probabilities to be of order unity (if there is one direction which has a much higher tunneling rate than the other directions, then $T_{ij}\approx1$), and the diagonal transition probabilities should be of order $1/J$, where $J$ is the dimensionality of the grid.  Since the probability to land on a penultimate vacuum is given by $Q_i$ or $P_i^{(0)}$ times the product of transition probabilities along the shortest path from the initial to final vacuum, one might conclude that vacua with axial flux configurations should be preferred.  But there is a competing effect - there are more vacua whose decay chains funnel into the diagonals, than along the axes.  In the watcher and causal patch measure, applied to a large BP landscape, this effect dominates.  One can think in terms of a flow of probability.  Even though an axial decay chain is more probable than a given diagonal chain, there are many more chains that start in the Planck shell and flow into a diagonal.  

We will shortly discuss another measure where this is not the case.  But we briefly mention, as pointed out in \cite{Bousso:2007er}, that if the dominant vacuum is chosen as an initial condition, the causal patch measure reproduces the results of the bubble abundance measure\footnote{In \cite{Bousso:2009mw} Bousso showed that his ``local'' causal patch measure is equivalent to his ``global'' light- cone time prescription \cite{Bousso:2009dm} when, and only when, the dominant vacuum is chosen as the initial condition for the causal patch prescription. The bubble abundance measure as implemented in \cite{SchwartzPerlov:2006hi}, is equivalent to the light-cone time prescription.}.  It would also reproduce the watcher measure with $Q_{dom}=1$.  Thus we see that, the $Q_i$ in the watcher measure, play the role of the initial probability distribution $P_i^{(0)}$ for the causal patch measure, even though these quantities have very different meanings.    

\subsection{Volume weighted-measures}

In \cite{Clifton:2007bn}, Clifton et al applied Linde's volume weighted measure \cite{Linde:2007nm} to a large BP landscape.  In this measure, probabilities of low energy vacua are weighted by the product of transition rates, $\kappa_{ij}$ (not transition probabilities ) along paths leading from initial to final vacua.  This leads to a very different prediction. Since transition rates vary exponentially, the ``axial'' paths are exponentially favored over the ``diagonal'' paths.  In the BP models studied so far, this effect by far overcompensates for the smaller hyper surface of initial high energy vacua which lie close to the axes.  Thus Clifton et al predicted that low energy vacua are far more likely to have one flux quantum several times higher than all the other quanta.  
 
\section{Conclusions} \label{conclusions}
We have applied the watcher measure to a small toy BP landscape.  The analytic results are
clear enough to allow us to generalize to large landscapes, and to compare with the results of other measures.   The watcher measure favors  the dynamical selection of low energy vacua with ``diagonal'' flux configurations, like the bubble abundance and causal patch measures; while volume weighted measures favor ``axial'' flux configurations.  

We find that the transition probabilities from crunching AdS vacua to dS vacua, $Q_j$, play a role that is analogous to the initial probability distribution, $P_i^{(0)}$, of the causal patch and volume-weighted measures \cite{Bousso:2007er,Clifton:2007bn} (see also \cite{Vanchurin:2006xp}).  If the physics underlying crunches can be well enough understood, so that we can calculate the crunching transition probabilities with confidence, then this measure will be strongly preferred over the initial condition dependent causal patch prescription, even though they make the same predictions.

\section{Acknowledgments} \label{appendix}

I would like to thank Alex Vilenkin for many useful discussions and comments.

\appendix
\section{Nucleation rates in the Bousso-Polchinski landscape}
In the BP model, we have a $J$-dimensional grid of vacua characterized
by the fluxes $F_a=n_a q_a$ and vacuum energy densities given by
Eq. (\ref{totalLambda}). Although the charges $q_a$ need not be very
small, $q_a/|\Lambda_{bare}|\sim 0.1~-~1$, the spectrum of values of $\Lambda_j$ may form a discretuum which is dense enough to explain why we can expect to find vacua with cosmological constants that lie in the anthropic range.

Transitions between neighboring vacua, which change one of the
integers $n_a$ by $\pm 1$ can occur through bubble
nucleation. The
bubbles are bounded by thin branes, whose tension $\tau_a$ is related
to their charge $q_a$ as \cite{BP,FMSW}
\beq
\tau_a^2 =q_a^2/2.
\label{tauj}
\eeq
We will not consider transitions for which $|\Delta n_a|>1$, or for which several $n_a$ are changed at once.

The bubble nucleation rate $\Gamma_{ij}$ per unit spacetime volume can
be expressed as \cite{CdL}
\be
\Gamma_{ij}=A_{ij} \exp^{-B_{ij}}
\label{Gamma}
\ee
with
\beq
B_{ij}=I_{ij}-S_j
\label{Bij}
\eeq
Here, $I_{ij}$ is the Coleman-DeLuccia instanton action and
\beq
S_j=-{8\pi^2\over{H_j^2}}
\label{Sj}
\eeq
is the background Euclidean action of de Sitter space.

In the case of a thin-wall bubble, which is appropriate for our
problem, the instanton action $I_{ij}$ has been calculated in
Refs.~\cite{CdL,BT}. It depends on the values of $\Lambda$ inside
and outside the bubble and on the brane tension $\tau$.

Let us first consider a bubble which changes the flux $a$ from $n_a$
to $n_a-1$ ($n_a>0$).  The resulting change in the cosmological constant is
given by
\be
|\Delta\Lambda_a|=(n_a-1/2)q_a^2,
\label{DeltaLambda}
\ee
and the exponent in the tunneling rate (\ref{Gamma}) can be
expressed as
\be
B_{a\downarrow} = B_{a\downarrow}^{flatspace} r(x,y).
\label{Bdown}
\ee
Here, $B_{a\downarrow}^{flatspace}$ is the flat space
bounce action,
\be
B_{a\downarrow}^{flatspace}= \frac{27
\pi^2}{2}\frac{\tau_a^4}{|\Delta \Lambda_a|^3}.
\ee
With the aid of
Eqs.~(\ref{tauj}),(\ref{DeltaLambda}) it can be expressed as
\be
B_{a\downarrow}^{flatspace}= \frac{27
\pi^2}{8}\frac{1}{(n_a-1/2)^3q_a^2}
\label{Bflat}
\ee

The gravitational correction factor $r(x,y)$ is given by
\cite{Parke}
\be
r(x,y) = \frac{2[(1+x
y)-(1+2xy+x^2)^{\frac{1}{2}}]}{x^2(y^2-1)(1+2xy+x^2)^{\frac{1}{2}}}
\label{gravfactor}
\ee
with the dimensionless parameters
\be
x\equiv
\frac{3q_a^2}{8|\Delta\Lambda_a|}=\frac{3}{8(n_a-1/2)}
\ee
and
\be
y\equiv \frac{2\Lambda}{|\Delta\Lambda_a|}-1, \label{y} \ee where
$\Lambda$ is the background value prior to nucleation.

The prefactors $A_{ij}$ in (\ref{Gamma}) can be estimated as
\beq
A_{ij} \sim 1.
\label{Aij}
\eeq

If the vacuum $n_a-1$ still has a positive energy density, then an
upward transition from $n_a -1$ to $n_a$ is also possible. The
corresponding transition rate is characterized by the same instanton
action and the same prefactor \cite{EWeinberg}, \beq I_{ij}=I_{ji},
~~~~~~~ A_{ij}=A_{ji}, \label{ijji} \eeq and it follows from
Eqs.(\ref{Gamma}), (\ref{Bij}) and (\ref{Hj}) that the upward and
downward nucleation rates are related by \be \Gamma_{ji} =
\Gamma_{ij} \exp\left[24 \pi^2
\left(\frac{1}{\Lambda_{j}}-\frac{1}{\Lambda_{i}}\right)\right].
\label{updown} \ee The exponential factor on the right-hand side of
(\ref{updown}) depends very strongly on the value of $\Lambda_{j}$.
The closer we are to $\Lambda_j=0$, the more suppressed are the
upward transitions $j\to i$ relative to the downward ones.

Eq.~(\ref{updown}) shows that the transition rate from $n_a$ up to
$n_{a+1}$ is suppressed relative to that from $n_{a+1}$ down to
$n_a$. It can also be shown that upward transitions from $n_a$ to
$n_{a+1}$ are similarly suppressed relative to the downward
transitions from $n_a$ to $n_{a-1}$.  

To develop some intuition for the dependence of the tunneling
exponent $B_{a\downarrow}$ on the parameters of the model $q_a$ and $n_a$, we shall
follow the qualitative discussion given in \cite{Bousso:2007er}.  We will focus on downward transitions.  Then ${{\partial B}\over {\partial n}}<0$.  This means that if we have a vacuum with a flux configuration that includes two unequal flux quanta ($n_i>n_j$, say), with approximately the same charge $q_i \approx q_j$, the larger flux $n_i$ is more likely to decay. The dependence on the charge is more complicated.  However, for most of the decay process, ${{\partial B}\over {\partial q}}>0$.  Thus, if two fluxes have equal quanta, then the one with the smallest charge is more likely to decay.  The dependence on the flux quanta is stronger than that on the size of the charges.

The main result is that as one follows a decay chain, from high to low energy vacua, fluxes with large $n_i$ (and small $q_i$), are the most likely to decay, and they do so earlier on in the chain.  By the time the penultimate vacua are reached, the dynamically selected vacua have flux configurations favoring evenly distributed low values of the quanta, close to ``diagonals''.


\begin{thebibliography}{99}


  
\bibitem{Linde:1993nz} 
  A.~D.~Linde and A.~Mezhlumian,
  Phys.\ Lett.\ B {\bf 307}, 25 (1993)
  [gr-qc/9304015].
  
  
\bibitem{Linde:1993xx} 
  A.~D.~Linde, D.~A.~Linde and A.~Mezhlumian,
  Phys.\ Rev.\ D {\bf 49}, 1783 (1994)
  [gr-qc/9306035].

\bibitem{GarciaBellido:1993wn} 
  J.~Garcia-Bellido, A.~D.~Linde and D.~A.~Linde,
  Phys.\ Rev.\ D {\bf 50}, 730 (1994)
  [astro-ph/9312039].
  
\bibitem{Vilenkin:1994ua} 
  A.~Vilenkin,
  Phys.\ Rev.\ Lett.\  {\bf 74}, 846 (1995)
  [gr-qc/9406010].

\bibitem{DeSimone:2008bq} 
  A.~De Simone, A.~H.~Guth, M.~P.~Salem and A.~Vilenkin,
  Phys.\ Rev.\ D {\bf 78}, 063520 (2008)
  [arXiv:0805.2173 [hep-th]].
  
\bibitem{Bousso:2008hz} 
  R.~Bousso, B.~Freivogel and I-S.~Yang,
  Phys.\ Rev.\ D {\bf 79}, 063513 (2009)
  [arXiv:0808.3770 [hep-th]].
  
\bibitem{DeSimone:2008if} 
  A.~De Simone, A.~H.~Guth, A.~D.~Linde, M.~Noorbala, M.~P.~Salem and A.~Vilenkin,
  Phys.\ Rev.\ D {\bf 82}, 063520 (2010)
  [arXiv:0808.3778 [hep-th]].
  
\bibitem{Garriga:2005av} 
  J.~Garriga, D.~Schwartz-Perlov, A.~Vilenkin and S.~Winitzki,
  JCAP {\bf 0601}, 017 (2006)
  [hep-th/0509184].
  
\bibitem{Bousso:2009dm} 
  R.~Bousso,
  Phys.\ Rev.\ D {\bf 79}, 123524 (2009)
  [arXiv:0901.4806 [hep-th]].
  
\bibitem{Vilenkin:2011yx} 
  A.~Vilenkin,
  JCAP {\bf 1106}, 032 (2011)
  [arXiv:1103.1132 [hep-th]].
  
\bibitem{Salem:2011mj} 
  M.~P.~Salem and A.~Vilenkin,
  Phys.\ Rev.\ D {\bf 84}, 123520 (2011)
  [arXiv:1107.4639 [hep-th]].
  
\bibitem{Linde:2008xf} 
  A.~D.~Linde, V.~Vanchurin and S.~Winitzki,
  JCAP {\bf 0901}, 031 (2009)
  [arXiv:0812.0005 [hep-th]].
  
 

\bibitem{VVW}
V.~Vanchurin, A.~Vilenkin, and S.~Winitzki, Phys.\ Rev.\ D {\bf 61},
083507 (2000).

\bibitem{Guth00}
A.~H.~Guth, Phys.\ Rept.\ {\bf 333}, 555 (2000).

\bibitem{Tegmark}
M.~Tegmark, JCAP {\bf 0504}, 001 (2005).



\bibitem{Bousso:2006ev} 
  R.~Bousso,
  Phys.\ Rev.\ Lett.\  {\bf 97}, 191302 (2006)
  [hep-th/0605263].
  
\bibitem{Bousso:2010zi} 
  R.~Bousso, B.~Freivogel, S.~Leichenauer and V.~Rosenhaus,
  Phys.\ Rev.\ Lett.\  {\bf 106}, 101301 (2011)
  [arXiv:1011.0714 [hep-th]].
  
  
\bibitem{Garriga:1997ef} 
  J.~Garriga and A.~Vilenkin,
  Phys.\ Rev.\ D {\bf 57}, 2230 (1998)
  [astro-ph/9707292].
  
\bibitem{Vanchurin:2006qp} 
  V.~Vanchurin and A.~Vilenkin,
  Phys.\ Rev.\ D {\bf 74}, 043520 (2006)
  [hep-th/0605015].
  
\bibitem{Vanchurin:2006xp} 
  V.~Vanchurin,
  Phys.\ Rev.\ D {\bf 75}, 023524 (2007)
  [hep-th/0612215].
  
\bibitem{Nomura:2011dt} 
  Y.~Nomura,
  JHEP {\bf 1111}, 063 (2011)
  [arXiv:1104.2324 [hep-th]].
  
  
\bibitem{Freivogel:2011eg} 
  B.~Freivogel,
  Class.\ Quant.\ Grav.\  {\bf 28}, 204007 (2011)
  [arXiv:1105.0244 [hep-th]].
  
  
  \bibitem{Watchers} 
  J.~Garriga and A.~Vilenkin,
  JCAP {\bf 1305}, 037 (2013)
  [arXiv:1210.7540 [hep-th]].
  

\bibitem{Garriga:2001ch} 
  J.~Garriga and A.~Vilenkin,
  Phys.\ Rev.\ D {\bf 64}, 043511 (2001)
  [gr-qc/0102010].

\bibitem{BP}
R. Bousso and J. Polchinski, JHEP {\bf 0006}, 006 (2000).


\bibitem{SchwartzPerlov:2006hi} 
  D.~Schwartz-Perlov and A.~Vilenkin,
  JCAP {\bf 0606}, 010 (2006)
  [hep-th/0601162].



\bibitem{Bousso:2007er} 
  R.~Bousso and I-S.~Yang,
  Phys.\ Rev.\ D {\bf 75}, 123520 (2007)
  [hep-th/0703206].
  
\bibitem{Linde:2007nm} 
  A.~D.~Linde,
  JCAP {\bf 0706}, 017 (2007)
  [arXiv:0705.1160 [hep-th]].
  
\bibitem{Clifton:2007bn} 
  T.~Clifton, S.~Shenker and N.~Sivanandam,
  JHEP {\bf 0709}, 034 (2007)
  [arXiv:0706.3201 [hep-th]].
  
\bibitem{Vilenkin:1984wp} 
  A.~Vilenkin,
  Phys.\ Rev.\ D {\bf 30}, 509 (1984).
  
\bibitem{Linde:1983mx} 
  A.~D.~Linde,
  Lett.\ Nuovo Cim.\  {\bf 39}, 401 (1984).
  
\bibitem{Bousso:2009mw} 
  R.~Bousso and I-S.~Yang,
  Phys.\ Rev.\ D {\bf 80}, 124024 (2009)
  [arXiv:0904.2386 [hep-th]].
  


\bibitem{FMSW}
J.L. Feng, J. March-Russell, S. Sethi and F. Wilczek, Nucl. Phys. {\bf
B602}, 307 (2001).

\bibitem{CdL}
S. Coleman and F. DeLuccia, Phys. Rev. {\bf D21}, 3305 (1980).


\bibitem{BT}
J. D. Brown and C. Teitelboim, Phys. Lett. \textbf{B195}, 177 (1987);
Nucl. Phys. \textbf{B297}, 787 (1988).

\bibitem{Parke}
S. Parke, ``Gravity and the decay of the false vacuum'',
{Phys}.\ Letters\ B {\bf 121} (1983) 313.


\bibitem{EWeinberg}
K.~M.~Lee and E.~J.~Weinberg, Phys.\ Rev.\ D {\bf 36}, 1088 (1987).

\end{thebibliography}
\end{document}